\pdfoutput=1

\documentclass[11pt,twoside,a4paper,cmspaper,final,collab]{cms-tdr}
\def\svnVersion{166212}\def\cmsCernNo{2012-037}\def\cmsCernDate{\today}\def\cmsMessage{Submitted to Physics Letters B}

\begin{document}\cmsNoteHeader{EXO-11-087}

\hyphenation{had-ron-i-za-tion}
\hyphenation{cal-or-i-me-ter}
\hyphenation{de-vices}
\RCS$Revision: 109287 $
\RCS$HeadURL: svn+ssh://svn.cern.ch/reps/tdr2/papers/EXO-11-087/trunk/EXO-11-087.tex $
\RCS$Id: EXO-11-087.tex 109287 2012-03-07 14:00:17Z antonius $
\newcommand{\ptg}{p_{\mathrm{T},\Pgm}}
\newlength\cmsFigWidth
\ifthenelse{\boolean{cms@external}}{\setlength\cmsFigWidth{0.85\columnwidth}}{\setlength\cmsFigWidth{0.45\textwidth}}
\ifthenelse{\boolean{cms@external}}{\providecommand{\cmsLeft}{top}}{\providecommand{\cmsLeft}{left}}
\ifthenelse{\boolean{cms@external}}{\providecommand{\cmsRight}{bottom}}{\providecommand{\cmsRight}{right}}
\cmsNoteHeader{EXO-11-087} 
\title{Search for large extra dimensions in dimuon and dielectron events in pp collisions at \texorpdfstring{$\sqrt{s} = 7$\TeV}{sqrt(s) = 7 TeV}}

\date{\today}

\abstract{
Results are presented from a search for large, extra spatial dimensions in events with either two isolated muons or two isolated electrons. The data are from proton-proton interactions at $\sqrt{s} = 7$\TeV collected with the CMS detector at the LHC. The size of the data sample corresponds to an integrated luminosity of approximately 2\fbinv. The observed dimuon and dielectron mass spectra are found to be consistent with standard-model expectations. Depending on the number of extra dimensions, the $95\%$ confidence level limits from the combined $\Pgm\Pgm$ and $\Pe\Pe$ channels range from ${M_{ \mathrm s} > 2.4\TeV}$ to ${M_{ \mathrm s} > 3.8\TeV}$, where $M_{ \mathrm s}$ characterizes the scale for the onset of quantum gravity.
}

\hypersetup{%
pdfauthor={CMS Collaboration},%
pdftitle={Search for large extra dimensions in dimuon and dielectron events in pp collisions at sqrt(s) = 7 TeV},%
pdfsubject={CMS},%
pdfkeywords={CMS, Physics, Exotica, Dimuon, Dielectron, ADD, Extra Dimensions, Gravitons, Field theories in dimensions other than four}}

\maketitle 

Models that extend the structure of space-time predict new phenomena beyond the standard model (SM) of particle physics. Additional spatial dimensions, essential for formulating quantum gravity in the context of string theory, have been proposed as a solution to the SM hierarchy problem \cite{arkani98:hlz,arkani99:hlz,Randall:1999ee}. In this letter, we present a search for events at large dimuon or dielectron invariant mass due to contributions from virtual graviton processes in the  Arkani-Hamed, Dimopoulos, Dvali (ADD) model \cite{arkani98:hlz,arkani99:hlz}.

The ADD model postulates the existence of compactified extra dimensions. Gravity is assumed to propagate in the entire higher-dimensional space, while particles of the SM are confined to a 3-dimensional slice of the multidimensional space. The resulting fundamental Planck energy scale $M_{ \mathrm D}$ in the ADD model can be reduced to significantly lower values than suggested by the apparent Planck mass $M_{ \mathrm{Pl} } \approx 1.2 \cdot 10^{19}$\GeV deduced for $3$ spatial dimensions. $M_{ \mathrm D}$ must be of the order of the scale of electroweak symmetry breaking to provide an explanation of the hierarchy problem. This scenario predicts phenomenological effects that might be observed in proton-proton collisions at the LHC. In this paper, we adopt the assumption \cite{han99:hlz,GRW99:extradim} that all extra dimensions are compactified on a torus of size $r$. In this case, $M_{ \mathrm D}$ is related to $ M_{ \mathrm{Pl}}$ through ${ M_{\mathrm{D} }^{n+2} = \left. {M_{ \mathrm{Pl}}^2 } \middle / {\left(8 \pi r^{n} \right)} \right. } $, where $n$ is the number of extra dimensions.

The graviton in this $(3+n)$-dimensional formulation can be equivalently expressed as a set of $3$-dimensional Kaluza--Klein (KK) modes \cite{Klein:1926tv} with different graviton masses. The coupling of the KK modes to the SM energy--momentum tensor leads to an effective theory with virtual-graviton exchange at leading order (LO) in perturbation theory. An ultraviolet (UV) cutoff $\Lambda$ must be introduced to avoid divergences in the summed contributions from all modes. A phenomenological consequence of the small mass separation between adjacent KK modes is an enhancement in the expected rate of dilepton events at large invariant masses that appears to be non-resonant. Depending on the details of the model, virtual-graviton effects can provide the dominant experimental ADD signature at high-energy colliders \cite{han99:hlz,GRW99:extradim}.

Several ways of parameterizing the LO differential cross sections are provided in the literature, including the Han, Lykken, Zhang (HLZ) \cite{han99:hlz} and the Giudice, Rattazzi, Wells (GRW) \cite{GRW99:extradim} conventions. In the GRW convention, the leading-order phenomenology for partonic centre-of-mass energies $\sqrt{\hat{s}} \ll \Lambda$ is described by a single parameter $\Lambda_{ \mathrm T}$, which does not depend on $\sqrt{\hat{s}}$ for $n \geq 3$.
The HLZ convention describes the phenomenology in terms of $n$ and a mass scale $M_{ \mathrm s}$, where $M_{ \mathrm s}$ is related to the selected UV cutoff and reflects the scale for the onset of quantum gravity. Typically, $M_{ \mathrm s}$ is expected to be of order $M_{ \mathrm D}$. The parameter $\Lambda_{ \mathrm T}$ can be related to the parameters in the HLZ convention \cite{Landsberg:2001proc}. The results of the analysis are interpreted in terms of both the HLZ and the GRW parameter conventions.

The effective theory breaks down at energy scales at which the underlying theory of
quantum gravity starts to affect the phenomenology. We assume that the range of validity is characterized by a value $\sqrt{\hat{s}_{\max}}$, roughly corresponding to the mass $M_{ \mathrm \max}$ of the lepton pairs emitted in the decay of the graviton. As no clear prediction for $\sqrt{\hat{s}_{\max}}$ can be made within the ADD model, and to take into account the requirement $\sqrt{\hat{s}_{\max}} \ll \Lambda$, most results in this paper are presented both for $M_{ \mathrm \max} = M_{ \mathrm s}$ and for a range of different values of $M_{ \mathrm \max}$.

Constraints on virtual-graviton signatures in the ADD model of extra dimensions have been obtained at HERA \cite{Adloff:2000dp,Chekanov:2003pw}, LEP \cite{Acciarri:1999bz,Acciarri:1999jy,Abreu:2000ap,Abreu:2000kh,Abbiendi:2002je,Abbiendi:1999wm}, and the Tevatron \cite{Abazov:2008as,Abazov:2005tk}. At the LHC, limits have been presented based on measurements of diphoton events \cite{Chatrchyan:2011jx,cmspaper11:ADDgammagamma,ATLAS:2011ab}.

CMS uses a right-handed coordinate system with axes labeled $x$, $y$, and $z$, and the origin at the center of the detector. The $z$-axis points along the direction of the anticlockwise beam. The azimuthal and polar angles are $\phi$ and $\theta$, with $\theta$ measured from the positive $z$-axis. The pseudorapidity $\eta$ is defined by $\eta = - \ln \tan (\theta / 2) $.

A main feature of the CMS apparatus is a superconducting solenoid of 6\unit{m} internal diameter, providing a magnetic field of 3.8\unit{T}. Located within the field volume are silicon pixel and strip inner trackers, an electromagnetic calorimeter (ECAL), and a hadronic calorimeter (HCAL).
The ECAL consists of lead-tungstate crystals covering pseudorapidities of $\vert \eta \vert< 1.5 $ (barrel) and $1.5 <\vert \eta \vert < 3.0$ (endcaps). The CMS muon detectors are embedded in the return yoke of the magnet. Muons are measured with detection planes using three different technologies: Drift Tubes, Cathode Strip Chambers, and Resistive Plate Chambers. The first stage of the CMS trigger system employs custom hardware and processes information from the calorimeters and the muon system. The event rate is further reduced by a computer farm using the event information from all detector systems. A detailed description of CMS can be found in Ref.~\cite{cms:2008zzk}.

This analysis uses data samples collected with the CMS detector in 2011, corresponding to an integrated luminosity \cite{pas:CMS-PAS-EWK-10-004} $\Lumi$ of $2.3 \pm 0.1\fbinv$ (dimuons) or $2.1 \pm 0.1\fbinv$ (dielectrons). The integrated luminosity for the dimuon channel is larger because the muon selection has less stringent requirements on the performance of the calorimeters during data-taking. The muon data sample was collected using a single-muon trigger with a transverse momentum ($\pt$) threshold which was varied between 15 and 40\GeV over the course of data-taking to allow for changes in instantaneous luminosity. The selection of electron events is based on a trigger requiring the presence of 2 electrons or photons with energy depositions $> 33$\GeV. Candidate events are required to have a reconstructed interaction vertex with $|z| < 24\unit{cm}$, and a radial distance $\sqrt{ x^2 + y ^2 } < 2\unit{cm}$. For events passing the complete selection requirements, the trigger efficiencies for signal and SM Drell--Yan (DY) events with large mass are $> 99 \%$, with an uncertainty of $< 1 \%$.

Muons with $|\eta| < 2.1$ and $\ptg> 45$\GeV are selected.
The candidates are required to be identified both in the outer muon system and the inner tracker, and the inner track must contain reconstructed energy deposits in at least 1 pixel layer and more than 10 strip-tracker layers. Muon candidates are required to have signals from at least two muon detector layers included in the reconstructed muon track.
Muon candidates satisfy the isolation requirement  $\sum{\pt^{i}} / \ptg< 0.1$, where the sum extends
over the momenta $\pt^{i}$ of all charged particle tracks (excluding the muon track) within a cone of size $\Delta R = \sqrt{ \left( \Delta \eta \right)^2 + \left( \Delta \phi \right)^2 } = 0.3$ around the
muon direction of flight. To reject backgrounds from cosmic-ray muons, we require a transverse impact parameter relative to the primary vertex of $ < 0.2$\unit{cm}, and an opening angle of $\alpha_{\Pgm\Pgm} < \pi - 0.02$ between the 2 muon momentum vectors. No charge requirement is applied to the muon pairs. However, all selected muon pairs of mass $> 450$\GeV are found to have opposite charges.
Events with 2 muons passing the selection criteria are accepted for analysis.

Electron candidates are reconstructed from energy depositions in the ECAL (superclusters) matched to
a track in the silicon tracker. ECAL superclusters are constructed from 1 or more clusters of energy depositions surrounding the crystal with the highest local energy deposition. An associated track is required to contain signals from at least 5 tracker layers. The track must be matched geometrically to the supercluster, and the spatial distribution of energy must be consistent with that expected for an electron. Only electron candidates with a ratio of energy depositions in the HCAL and ECAL below $0.05$ are considered. To minimize the contamination from jets, electron candidates are required to be isolated. Candidates with a sum of transverse track momenta $\ge 5$\GeV within ${0.04 < \Delta R < 0.3}$ around the candidate track are rejected. In the ECAL and inner HCAL layer, the deposited transverse energy $\ET$ in a cone $\Delta R = 0.3$ around the electron candidate (excluding the transverse energy $E_{{\mathrm T},\Pe}$ of the electron) must be ${< 2\GeV + 0.03 \times E_{{\mathrm T},\Pe}}$ for the barrel, or $< 2.5\GeV ({<}2.5
\GeV + 0.03 \times (E_{{\mathrm T},e} -50\GeV))$ for the endcaps if ${E_{{\mathrm T},e} < 50}\GeV ({E_{{\mathrm T},e} \geq 50}\GeV)$. Additionally, the $E_{{\mathrm T}}$ deposition in the outer HCAL layer within ${0.15 < \Delta R < 0.3}$ around the electron position is restricted to $< 0.05$\GeV. Selected events must contain 2 electrons of opposite charge,
each with transverse energy ${\ET > 35}$\GeV (in the barrel region), or with ${\ET > 40}$\GeV (in the endcaps). The explicit charge requirement is found to have negligible influence on the presented results. Events in which both electrons are reconstructed in the endcaps are not used in the analysis, since electrons from the ADD signal would on average be produced at smaller values of $\eta$ than the SM backgrounds.

The search is performed with a set of events that contains either electron or muon pairs above a mass value $M_{\mathrm{min}}$. The lower bound of the signal region is chosen to maximize the expected upper limits of the ADD model parameter $\Lambda_{ \mathrm T}$ in each lepton channel.
The optimum value of ${ M_{\mathrm{min}} }$ is found to be $1.1$\TeV for both the dimuon and the dielectron channel, based on simulation studies.

In both search channels, the \PYTHIA8.142 \cite{Skands08:Pythia8,Ask09:ADDpythia8} event generator with the \textsc{mstw}08 \cite{Martin:2009MSTW} parton distribution function (PDF) set is used to simulate the expected signal. Interference terms between the standard model DY process and the virtual graviton are taken into account in the evaluation of the signal cross sections. Simulated events for both signal and SM backgrounds are passed through a detailed detector simulation based on \GEANTfour \cite{Agostinelli:2002hh}, using a realistic CMS alignment scenario, and the same reconstruction chain as data.

In this analysis, the SM DY process is the dominant background. In the dimuon channel, we use the \MCATNLO \cite{Frixione02:MCatNLO,Frixione10:MCatNLO} event generator with the \textsc{cteq}6.6 \cite{Nadolsky:2008zw} PDF set to simulate the DY background. The parton level events from \MCATNLO are passed to \HERWIG6 \cite{Corcella:2000Herwig} for the simulation of the QCD parton shower and hadronization, \PHOTOS \cite{Golonka:2005Photos} for the simulation of the electroweak (EW) parton shower, and \textsc{Jimmy} \cite{Butterworth:1996Jimmy} for the simulation of multiple parton interactions. The simulated reconstruction efficiencies in the chosen region of acceptance, including all selection criteria, are found to be $90\% \pm 3\%$ for the high-mass DY dimuon background and $90\% \pm 4\%$ for the ADD dimuon signal.

Mass-dependent corrections~\cite{Ballossini08:DrellYanNLO} beyond the QCD next-to-leading order (NLO) predictions implemented in \MCATNLO are studied to improve the SM DY estimate in the dimuon channel. EW NLO effects are evaluated by comparing {\sc horace} \cite{Carloni07:Horace} NLO predictions interfaced to \HERWIG6 with \textsc{horace} LO predictions interfaced to \HERWIG6 and \PHOTOS. In this procedure, \PHOTOS corrections are applied to the LO results to account for radiation effects as these corrections are also included in the DY simulation based on \MCATNLO. The effect of electroweak NLO corrections is found to be smaller than the QCD NLO contribution and of opposite sign. The estimated correction factor for the DY background beyond $1.1\TeV$ is $ \approx 0.90 \pm 0.06$. Next-to-next-to-leading order (NNLO) QCD corrections are obtained using higher-order calculations from \textsc{fewz} \cite{petriello10:FEWZ}. The corresponding multiplicative correction factor for the DY background in the signal region is estimated to be $1.03 \pm 0.03$. For the purpose of setting limits, both the EW NLO and QCD NNLO corrections are applied to the DY background prediction obtained from \MCATNLO.

The DY background in the dielectron channel is simulated with \PYTHIA6 \cite{Sjostrand:2006za} and normalized according to the observed data in the range 60--120\GeV around the \Z resonance.  As in the dimuon channel, electroweak NLO corrections at large mass are studied with \textsc{horace}. The estimated correction factor for the DY background beyond 1.1\TeV is found to be $0.92 \pm 0.06$. The simulated reconstruction efficiency for the high-mass DY dielectron background and the ADD dielectron signal in the selected acceptance range, including all selection criteria, is found to be $84\% \pm 3\%$.

The parton distribution functions have a strong impact on the SM DY background in both search channels. The PDF uncertainties for the DY process are evaluated by comparing results from the \textsc{cteq}6.6 \cite{Nadolsky:2008zw}, \textsc{mstw}08 \cite{Martin:2009MSTW}, and \textsc{nnpdf}21 \cite{Ball:2009NNPDF} PDF groups. This procedure follows the recommendations of the PDF4LHC working group \cite{Alekhin:2011PDF4LHC}. The uncertainties are defined by constructing an envelope that embraces the three separate PDF sets from the respective groups, together with their individual associated uncertainties. Within each group, PDF reweighting \cite{Tricoli:2005nx} is used to evaluate the respective uncertainties. Additional uncertainties from the dependence on the strong coupling constant $\alpha_s$ are estimated with \textsc{mstw}08 PDF sets. The resulting uncertainty on the integrated SM DY distribution for masses above $1.1$\TeV, from all uncertainties related to the choice of PDF, is estimated to be $13\%$.

\begin{figure}[htbp]
  \centering
    \includegraphics[width=\cmsFigWidth]{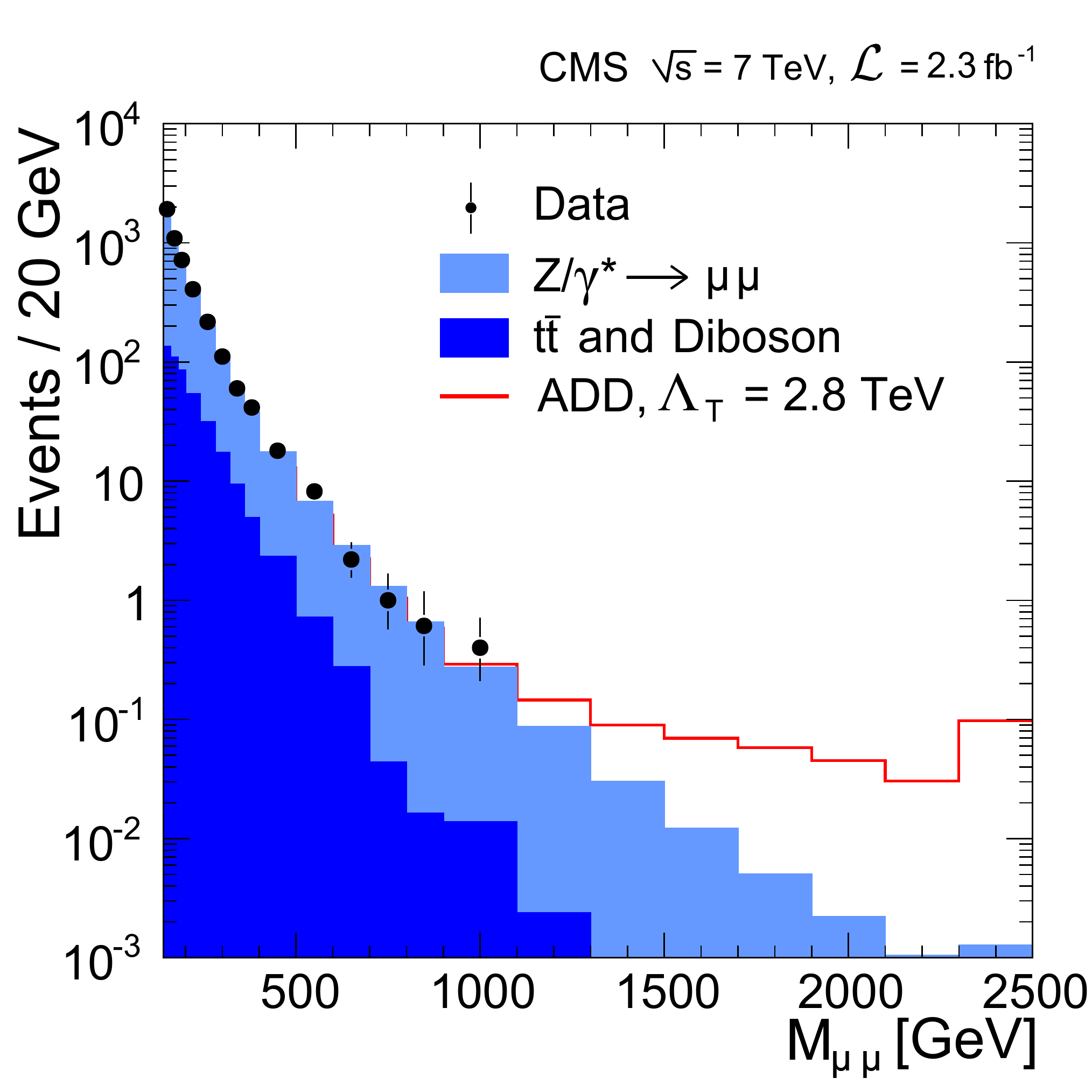}
    \includegraphics[width=\cmsFigWidth]{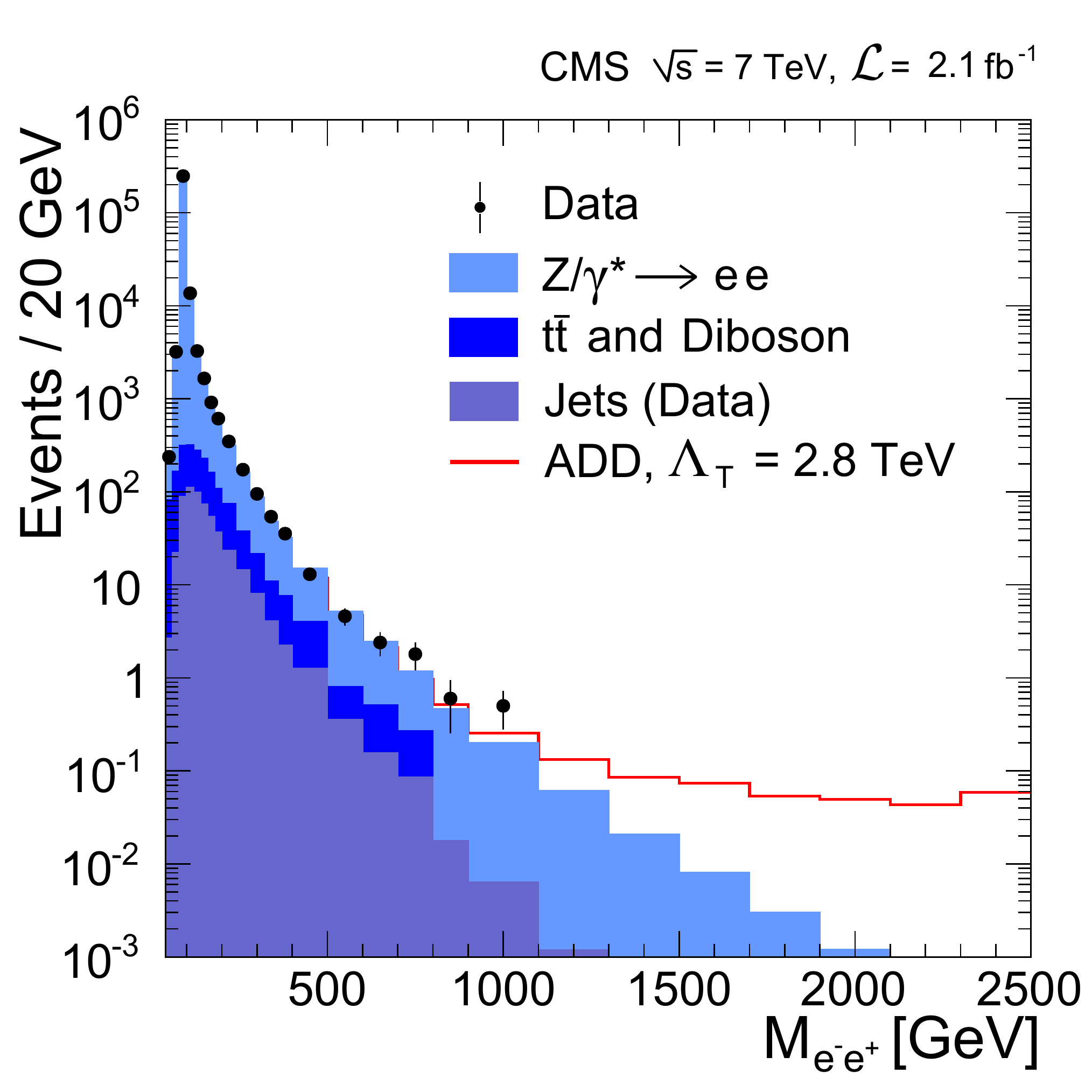}
  \caption{Dimuon (\cmsLeft) and dielectron (\cmsRight) invariant mass spectra compared with the stacked SM predictions and an added simulated
    ADD signal with $\Lambda_{ \mathrm T} = 2.8$\TeV (ADD K-factor 1.0, no signal truncation). The highest-mass bins contain all contributions above $2.3$\TeV. The error bars reflect the statistical uncertainty.}
  \label{fig:data_dist}
\end{figure}

Contributions from $\ttbar$, $\cPqt\PW$, and EW vector boson pair production to the dimuon and dielectron mass spectrum are estimated by using simulations with \MADGRAPH \cite{Alwall:2011uj} and \PYTHIA6.  The background contributions are cross-checked with a control sample dominated by these processes, including events with an electron and a muon passing requirements similar to those used for the signal leptons. Taking into account the differences in the acceptance and efficiencies between muons and electrons, the ratios between the expected $\Pe\Pe$, $\Pgm\Pgm$, and $\Pe\Pgm$ backgrounds from the $\ttbar$, $\cPqt\PW$, and diboson contributions in the SM are well understood from lepton universality. The measured $\Pe\Pgm$ mass spectrum is found to be well reproduced by the simulations. The agreement has been confirmed up to masses of $\approx 500$\GeV, above which the statistical uncertainties on the $\Pe\Pgm$ spectrum become large.

Background contributions at large dimuon mass from multijet processes and cosmic-ray muons are negligible for our event selection requirements.

In addition to those backgrounds that are common with the dimuon channel, the dielectron channel receives background contributions from multijet events with 2 jets that pass the electron selection and $\PW+\text{jets}$ events with 1 jet passing electron selection. Events of the type ${\Pgg+\text{jets}}$, where the photon converts to $\Pep\Pem$ and both the photon and a jet are reconstructed as electrons that pass selection, are also considered. The rate for jets to be reconstructed as electrons is determined from a control sample of events selected by a single-electromagnetic-cluster trigger with a lower threshold. The electron selection criteria, including the isolation requirements, are relaxed to define electron candidates in this sample. Events are required to have no more than 1 such reconstructed electron to suppress the contribution from the DY process. Residual contributions from ${\PW+\text{jets}}$ and ${\Pgg+\text{jets}}$ events in the control sample are estimated from simulation. The estimated probability for an electron candidate to pass the full set of electron selection criteria is then used to weight the events that have 2 such candidates passing the double-electromagnetic-cluster trigger.

Both for dimuon and dielectron events, the contributions from non-DY processes sum up to less than $10\%$ of the expected backgound in the signal region.

Using \Z-candidate events, detailed studies are performed of possible differences in the electron and muon reconstruction efficiencies of simulated events and data \cite{:2011nx}. No statistically significant deviations between data and simulations are found, indicating that the simulated lepton reconstruction efficiencies are reliable. In both channels, uncertainties on the simulated acceptance and reconstruction efficiencies at large dilepton mass are included in the statistical evaluation of the result. Uncertainties related to momentum reconstruction of muons and energy estimation of electrons are also taken into account.

The systematic uncertainties for the integrated dimuon and dielectron backgrounds in the signal region are summarized in Table \ref{tab:overview_systematics}.
With the exception of the uncertainty on the integrated luminosity, which is treated as fully correlated between the 2 channels, they are assumed to be independent.

Figure~\ref{fig:data_dist} shows the observed and expected mass distributions in the 2 search channels as a function of dilepton mass. Measurements and SM predictions are found to be in agreement within statistical and systematic uncertainties. In both channels, no significant excess of events is observed in the high-mass region, and no events are found in the signal region. The corresponding SM background expectation in the signal region is $1.0 \pm 0.2$ events in the dielectron channel and $1.3 \pm 0.2$ events in the dimuon channel. The observed number of events $N_{\text{obs}}$ and the SM expectation are in agreement in several control regions, as shown in in Table \ref{tab:dataMCscalefactors}.

\begin{table*}[htbp]
\centering
\caption{Summary of systematic uncertainties for the integrated
dimuon and dielectron invariant mass spectra in the signal regions.}
\begin{tabular}{|l|c|c|}
\hline
Systematic uncertainty  & Uncertainty & Uncertainty \\
                        & on signal $( \% )$ & on background $( \% )$ \\
\hline
Integrated luminosity                         &  $4.5$ & $4.5$ \\
Trigger and reconstruction efficiency                     & $4 $ $(\Pgm\Pgm)$, $3 $ $(\Pe\Pe)$ & $3 $ $(\Pgm\Pgm)$, $3 $ $(\Pe\Pe)$ \\
Muon momentum resolution                      & $1 $ & $5 $  \\
Electron energy scale                         &  $1-3$ & $1-3$ \\
Drell--Yan PDF uncertainties                  & --- & $13 $  \\
Drell--Yan higher order corrections           & --- & $10 $  \\
\hline
\end{tabular}
\label{tab:overview_systematics}
\end{table*}

\begin{table*}[htbp]
  \centering
  \caption{Comparison of the observed and expected number of events in control and signal regions for the dimuon and dielectron mass distributions. Expected signal contributions are shown for ${\Lambda_{ \mathrm T} = 2.8}$\TeV (ADD K-factor 1.0, signal truncation at $M_{\mathrm{max}} = \Lambda_{ \mathrm T}$). }
      \small{
      \begin{tabular}{|c|r|r@{$\pm$}l|r|}
      \hline
      \multicolumn{5}{|c|}{ \parbox[0pt][2em][c]{0cm}{} $\Pgm\Pgm$, $\Lumi = 2.3\fbinv$} \\
      \hline
      \multicolumn{1}{|l|}{Mass}                         & $N_{\text{obs}} $ & \multicolumn{2}{c|}{Background}  & Signal exp. \\
      \multicolumn{1}{|l|}{region $ [\TeVns] $}    &                     & \multicolumn{2}{c|}{expectation} & \scriptsize{${\Lambda_{ \mathrm T} = 2.8}$\TeV} \\
      \hline
      \hline
      \multicolumn{5}{|l|}{Control regions} \\
      \hline
      \hline
      0.14--0.20 & $3723$ & ~ $3690$ & $300$ & - \\
      \hline
      0.20--0.40 & $1674$ & ~ $1605$  & $160$ & - \\
      \hline
      0.40--0.60 & $131$ & ~ $122$ & $13$ & - \\
      \hline
      0.60--0.80 & $16$  & ~ $21$ & $3$ & - \\
      \hline
      0.80--1.10 & $7$  & ~ $6$ & $1$ & $0.8$ \\
      \hline
      \hline
      \multicolumn{5}{|l|}{Signal region} \\
      \hline
      \hline
      $>$ 1.10 & $0$  & ~ $1.3$ & $0.2$ & $3.2$ \\
      \hline
      \end{tabular}
      \begin{tabular}{|c|r|r@{$\pm$}l|r|}
      \hline
      \multicolumn{5}{|c|}{ \parbox[0pt][2em][c]{0cm}{} $\Pe\Pe$, $\Lumi = 2.1\fbinv$} \\
      \hline
      \multicolumn{1}{|l|}{Mass}                         & $N_{\text{obs}} $ & \multicolumn{2}{c|}{Background}  & Signal exp. \\
      \multicolumn{1}{|l|}{region $ [\TeVns] $}    &                     & \multicolumn{2}{c|}{expectation} & \scriptsize{${\Lambda_{ \mathrm T} = 2.8}$\TeV} \\
      \hline
      \hline
      \multicolumn{5}{|l|}{Control regions} \\
      \hline
      \hline
      0.12--0.20 & $6592$ & ~ $6598$ & $530$ & - \\
      \hline
      0.20--0.40 & $1413$ & ~ $1301$ & $120$ & - \\
      \hline
      0.40--0.60 & $88$ & ~ $103$ & $11$ & - \\
      \hline
      0.60--0.80 & $21$ & ~ $18$ & $3$ & - \\
      \hline
      0.80--1.10 & $8$  & ~ $5$ & $1$ & $0.6$ \\
      \hline
      \hline
      \multicolumn{5}{|l|}{Signal region} \\
      \hline
      \hline
      $>$ 1.10 & $0$  & ~ $1.0$ & $0.2$ & $2.7$ \\
      \hline
      \end{tabular}
      }
  \label{tab:dataMCscalefactors}
\end{table*}

For the statistical evaluation of the measurements, we count the observed events in the signal region. For each channel, the probability of observing $N_{\text{obs}}$ events in the signal region is given by a Poisson distribution. The statistical model for the Poisson means includes parameters that are used to describe the influence of the systematic uncertainties listed in Table \ref{tab:overview_systematics} on the expected signal and background events.
Limits on the cross sections for signals in the regions of acceptance are calculated with a $\mbox{CL}_{\mbox{s}}$ approach \cite{Read:2000ru}. The applied test statistic is a one-sided profile likelihood ratio \cite{Cowan:2010js} corresponding to the selected models. The systematic uncertainties are included in the statistical evaluation by extending the likelihood with additional probability density functions that parameterize the respective uncertainties. The exclusion threshold is set to $\mbox{CL}_{\mbox{s}} = 0.05$ ( $> 95 \%$ confidence).

The RooStats \cite{RooStats:Ref1} software for statistical data analysis is used for the numerical evaluation of the $\mbox{CL}_{\mathrm{s}}$ limits.
At $95\%$ confidence level (CL), we exclude signal cross sections $\sigma_{\mathrm{s}}$ above 1.2\unit{fb} (1.8\unit{fb} expected) in the dimuon channel and 1.6\unit{fb} (2.3\unit{fb} expected) in the dielectron channel. The combined upper limit at $95\%$ CL on the signal cross section in both channels ${\sigma_{\mathrm{s,} \Pgm \Pgm + \Pe\Pe}}$ is found to be $1.4$\unit{fb}, while the expected limit is $2.2$\unit{fb}.

\begin{figure}[htbp]
  \centering
    \includegraphics[width=\cmsFigWidth]{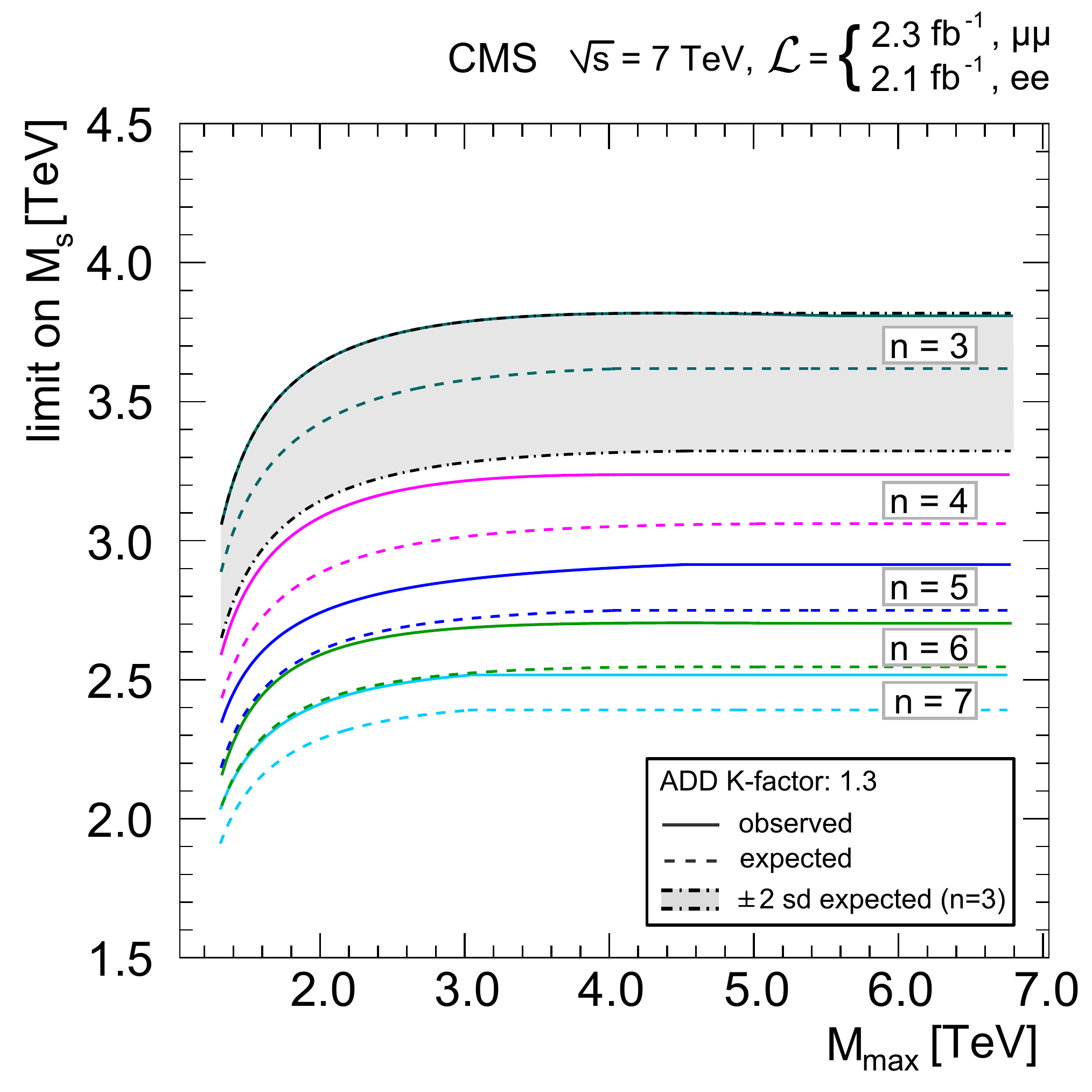}
  \caption{Observed and expected $95\%$ CL lower limits on $M_{ \mathrm s}$, obtained by combining the $\Pgm\Pgm$ and $\Pe\Pe$ results, for different numbers of extra dimensions $n$, applying a signal K-factor of $1.3$. A confidence interval for the expected limit corresponding to 2 standard deviations (sd) is shown for the case $n = 3$.}
  \label{fig:comb_limits_Ms_graphs_profile}
\end{figure}

The observed limits on $\sigma_{\mathrm{s}}$ are translated into exclusion limits on the ADD parameters. To account for interference effects, the expected signal contribution for a particular choice of model parameters is evaluated by subtracting the SM DY cross section at LO from the cross section with the ADD LO contributions. Limits are based either on the leading-order ADD scenario without higher-order corrections or on an assumed higher-order correction factor (K-factor) of $1.3$ for the ADD signal contributions.  Based on studies of QCD NLO corrections to dilepton processes in the ADD model \cite{Kumar:2006id,Mathews:2004xp}, the K-factor of $1.3$ corresponds to a conservative choice. Figure~\ref{fig:comb_limits_Ms_graphs_profile} shows the limits for the HLZ convention
for different ranges of validity of the model, assuming a K-factor of $1.3$ and no signal contribution beyond the cutoff $M_{\text{max}}$. Table \ref{tab:overview_results} summarizes the limits on the GRW and HLZ parameters for truncation of the signal at ${M_{ \text{max} } = \Lambda_{ \mathrm T}}$ or ${M_{ \text{max} } = M_{ \mathrm s}}$. Results are also given for an evaluation of limits separately for dimuon or dielectron measurements.
Including our recently published results on diphoton events \cite{cmspaper11:ADDgammagamma}, which have comparable sensitivity, improves the observed combined limits on $M_{ \mathrm s}$ presented in Table \ref{tab:overview_results} by $0.1$~$(0.3)$\TeV for $n=2$ and $0.1$~$(0.1)$\TeV for $n \geq 3$ without (with) K-factors for the signal contributions.

\begin{table*}[htbp]
\centering
\caption{Observed lower limits in \TeV at $95\%$ CL within GRW and HLZ conventions for truncation at $M_{\mathrm{max}} = \Lambda_{ \mathrm T}$ (GRW) or $M_{\mathrm{max}} = M_{ \mathrm s}$ (HLZ).}
\begin{tabular}{|c|c|cccccc|}
\hline
 ADD K-factor & $\Lambda_{ \mathrm T} \, [\TeVns]$ (GRW) & \multicolumn{6}{c|}{ $M_{ \mathrm s} \, [\TeVns]$ (HLZ) } \\
  & & $n = 2$ & $n = 3$ & $n = 4$ & $n = 5$ & $n = 6$ & $n = 7$ \\
\hline
\hline
\multicolumn{3}{|r@{$<$}}{$\Pgm \Pgm$, $\sigma_{\mathrm{s,} \Pgm \Pgm }$} & \multicolumn{5}{l|}{$1.2$\unit{fb} ($1.8$\unit{fb} expected) at $95\%$ CL } \\
\hline
\hline
1.0 & 2.8 & 3.0 & 3.4 & 2.8 & 2.5 & 2.3 & 2.2 \\
\hline
1.3 &  3.0 & 3.2 & 3.5 & 3.0 & 2.7 & 2.4 & 2.3 \\
\hline
\hline
\multicolumn{3}{|r@{$<$}}{$\Pe\Pe$, $\sigma_{\mathrm{s, \Pe\Pe}}$} & \multicolumn{5}{l|}{$1.6$\unit{fb} ($2.3$\unit{fb} expected) at $95\%$ CL } \\
\hline
\hline
1.0 &  2.8 & 2.9 & 3.3 & 2.8 & 2.5 & 2.3 & 2.2 \\
\hline
1.3 &  2.9 & 3.1 & 3.4 & 2.9 & 2.5 & 2.4 & 2.2 \\
\hline
\hline
\multicolumn{3}{|r@{$<$}}{$\Pgm \Pgm$ and $\Pe\Pe$, $\sigma_{\mathrm{s,} \Pgm \Pgm + \Pe\Pe}$} & \multicolumn{5}{l|}{$1.4$\unit{fb} ($2.2$\unit{fb} expected) at $95\%$ CL } \\
\hline
\hline
1.0 &  3.1 & 3.7 & 3.7 & 3.1 & 2.8 & 2.5 & 2.4 \\
\hline
1.3 &  3.2 & 3.8 & 3.8 & 3.2 & 2.9 & 2.7 & 2.5 \\
\hline
\end{tabular}
\label{tab:overview_results}
\end{table*}

In summary, a search for the effects of large extra dimensions in dimuon
and dielectron invariant mass spectra using the CMS detector at the LHC has been presented. The results are found to be in agreement with SM predictions, and no significant excess of events is observed at large values of dimuon or dielectron mass. In the signal region of dilepton masses above $1.1$\TeV, no events are found.
Our results extend the limits on ADD models based on the analysis of dilepton signatures. The combination with diphoton results provides the most stringent limits on graviton decay in the ADD framework to date.

We thank M.C. Kumar, P. Mathews, and V. Ravindran for useful discussions on QCD NLO corrections in the ADD model. We wish to congratulate our colleagues in the CERN accelerator departments for the excellent performance of the LHC machine. We thank the technical and administrative staff at CERN and other CMS institutes, and acknowledge support from: FMSR (Austria); FNRS and FWO (Belgium); CNPq, CAPES, FAPERJ, and FAPESP (Brazil); MES (Bulgaria); CERN; CAS, MoST, and NSFC (China); COLCIENCIAS (Colombia); MSES (Croatia); RPF (Cyprus); Academy of Sciences and NICPB (Estonia); Academy of Finland, MEC, and HIP (Finland); CEA and CNRS/IN2P3 (France); BMBF, DFG, and HGF (Germany); GSRT (Greece); OTKA and NKTH (Hungary); DAE and DST (India); IPM (Iran); SFI (Ireland); INFN (Italy); NRF and WCU (Korea); LAS (Lithuania); CINVESTAV, CONACYT, SEP, and UASLP-FAI (Mexico); MSI (New Zealand); PAEC (Pakistan); MSHE and NSC (Poland); FCT (Portugal); JINR (Armenia, Belarus, Georgia, Ukraine, Uzbekistan); MON, RosAtom, RAS and RFBR (Russia); MSTD (Serbia); MICINN and CPAN (Spain); Swiss Funding Agencies (Switzerland); NSC (Taipei); TUBITAK and TAEK (Turkey); STFC (United Kingdom); DOE and NSF (USA).
Individuals have received support from the Marie-Curie programme and the European Research Council (European Union); the Leventis Foundation; the A. P. Sloan Foundation; the Alexander von Humboldt Foundation; the Belgian Federal Science Policy Office; the Fonds pour la Formation \`a la Recherche dans l'Industrie et dans l'Agriculture (FRIA-Belgium); the Agentschap voor Innovatie door Wetenschap en Technologie (IWT-Belgium); the Council of Science and Industrial Research, India; and the HOMING PLUS programme of Foundation for Polish Science, cofinanced from European Union, Regional Development Fund.

\clearpage

\bibliography{auto_generated}   

\providecommand{\href}[2]{#2}\begingroup\raggedright\begin{thebibliography}{10}%
\makeatletter
\providecommand{\hrefCMSnoop }[0]{\@secondoftwo}%
\makeatother
\providecommand{\doi}{\texttt{doi:}\begingroup \urlstyle{tt}\Url}

\bibitem{arkani98:hlz}
\hrefCMSnoop {} {N.~Arkani-Hamed, S.~Dimopoulos, and G.~Dvali, ``The hierarchy
  problem and new dimensions at a millimeter'',} \textit{ Phys. Lett. B}
  \textbf{ 429} (1998) 263,
  \href{http://dx.doi.org/10.1016/S0370-2693(98)00466-3}{\doi{10.1016/S0370-2693(98)00466-3}}.

\bibitem{arkani99:hlz}
\hrefCMSnoop {} {N.~Arkani-Hamed, S.~Dimopoulos, and G.~Dvali, ``Phenomenology,
  astrophysics and cosmology of theories with sub-millimeter dimensions and
  {TeV} scale quantum gravity'',} \textit{ Phys. Rev. D} \textbf{ 59} (1999)
  086004,
  \href{http://dx.doi.org/10.1103/PhysRevD.59.086004}{\doi{10.1103/PhysRevD.59.086004}}.

\bibitem{Randall:1999ee}
\hrefCMSnoop {} {L.~Randall and R.~Sundrum, ``A Large Mass Hierarchy from a
  Small Extra Dimension'',} \textit{ Phys. Rev. Lett.} \textbf{ 83} (1999)
  3370,
  \href{http://dx.doi.org/10.1103/PhysRevLett.83.3370}{\doi{10.1103/PhysRevLett.83.3370}}.

\bibitem{han99:hlz}
\hrefCMSnoop {} {T.~Han, J.~D. Lykken, and R.~Zhang, ``On {K}aluza-{K}lein
  States from Large Extra Dimensions'',} \textit{ Phys. Rev. D} \textbf{ 59}
  (1999) 105006,
  \href{http://dx.doi.org/10.1103/PhysRevD.59.105006}{\doi{10.1103/PhysRevD.59.105006}}.

\bibitem{GRW99:extradim}
\hrefCMSnoop {} {G.~F. Giudice, R.~Rattazzi, and J.~D. Wells, ``Quantum gravity
  and extra dimensions at {H}igh-{E}nergy {C}olliders'',} \textit{ Nucl. Phys.
  B} \textbf{ 544} (1999) 3,
  \href{http://dx.doi.org/10.1016/S0550-3213(99)00044-9}{\doi{10.1016/S0550-3213(99)00044-9}}.

\bibitem{Klein:1926tv}
\hrefCMSnoop {} {O.~Klein, ``{Quantum theory and five-dimensional theory of
  relativity}'',} \textit{ Z. Phys.} \textbf{ 37} (1926) 895,
  \href{http://dx.doi.org/10.1007/BF01397481}{\doi{10.1007/BF01397481}}.

\bibitem{Landsberg:2001proc}
\href {http://cdsweb.cern.ch/record/476781} {G.~Landsberg, ``Extra Dimensions
  and More\ldots'',} in \textit{ 36th Rencontre de Moriond}.
\newblock Les Arcs, France, 2001.

\bibitem{Adloff:2000dp}
\hrefCMSnoop {} {{ H1} Collaboration, ``Search for Compositeness, Leptoquarks
  and Large Extra Dimensions in $eq$ Contact Interactions at {HERA}'',}
  \textit{ Phys. Lett. B} \textbf{ 479} (2000) 358,
  \href{http://dx.doi.org/10.1016/S0370-2693(00)00332-4}{\doi{10.1016/S0370-2693(00)00332-4}}.

\bibitem{Chekanov:2003pw}
\hrefCMSnoop {} {{ ZEUS} Collaboration, ``Search for contact interactions,
  large extra dimensions and finite quark radius in $ep$ collisions at
  {HERA}'',} \textit{ Phys. Lett. B} \textbf{ 591} (2004) 23,
  \href{http://dx.doi.org/10.1016/j.physletb.2004.03.081}{\doi{10.1016/j.physletb.2004.03.081}}.

\bibitem{Acciarri:1999bz}
\hrefCMSnoop {} {{ L3} Collaboration, ``Search for extra dimensions in boson
  and fermion pair production in {$\Pep\Pem$} interactions at {LEP}'',}
  \textit{ Phys. Lett. B} \textbf{ 470} (1999) 281,
  \href{http://dx.doi.org/10.1016/S0370-2693(99)01310-6}{\doi{10.1016/S0370-2693(99)01310-6}}.

\bibitem{Acciarri:1999jy}
\hrefCMSnoop {} {{ L3} Collaboration, ``Search for low scale gravity effects in
  {$\Pep\Pem$} collisions at {LEP}'',} \textit{ Phys. Lett. B} \textbf{ 464}
  (1999) 135,
  \href{http://dx.doi.org/10.1016/S0370-2693(99)01011-4}{\doi{10.1016/S0370-2693(99)01011-4}}.

\bibitem{Abreu:2000ap}
\hrefCMSnoop {} {{ DELPHI} Collaboration, ``Measurement and interpretation of
  fermion-pair production at {LEP} energies of 183 and 189 {GeV}'',} \textit{
  Phys. Lett. B} \textbf{ 485} (2000) 45,
  \href{http://dx.doi.org/10.1016/S0370-2693(00)00675-4}{\doi{10.1016/S0370-2693(00)00675-4}}.

\bibitem{Abreu:2000kh}
\hrefCMSnoop {} {{ DELPHI} Collaboration, ``Determination of the e+ e-
  $\rightarrow$ $\gamma$ $\gamma$ ($\gamma$) Cross-Section at Centre-of-Mass
  Energies Ranging from 189 {GeV} to 202 {GeV}'',} \textit{ Phys. Lett. B}
  \textbf{ 491} (2000) 67,
  \href{http://dx.doi.org/10.1016/S0370-2693(00)01013-3}{\doi{10.1016/S0370-2693(00)01013-3}}.

\bibitem{Abbiendi:2002je}
\hrefCMSnoop {} {{ OPAL} Collaboration, ``Multi-photon production in e$^+$
  e$^-$ collisions at {$\sqrt{s} = 181-209\GeV$}'',} \textit{ Eur. Phys. J. C}
  \textbf{ 26} (2003) 331,
  \href{http://dx.doi.org/10.1140/epjc/s2002-01074-5}{\doi{10.1140/epjc/s2002-01074-5}}.

\bibitem{Abbiendi:1999wm}
\hrefCMSnoop {} {{ OPAL} Collaboration, ``Tests of the standard model and
  constraints on new physics from measurements of fermion pair production at
  189 {GeV} at {LEP}'',} \textit{ Eur. Phys. J. C} \textbf{ 13} (2000) 553,
  \href{http://dx.doi.org/10.1007/s100520050718}{\doi{10.1007/s100520050718}}.

\bibitem{Abazov:2008as}
\hrefCMSnoop {} {{ D0} Collaboration, ``{Search for Large Extra Spatial
  Dimensions in the Dielectron and Diphoton Channels in p anti-p Collisions at
  $\sqrt{s} = 1.96 \, \mbox{\TeV}$}'',} \textit{ Phys. Rev. Lett.} \textbf{
  102} (2009) 051601,
  \href{http://dx.doi.org/10.1103/PhysRevLett.102.051601}{\doi{10.1103/PhysRevLett.102.051601}}.

\bibitem{Abazov:2005tk}
\hrefCMSnoop {} {{ D0} Collaboration, ``{Search for Large Extra Spatial
  Dimensions in Dimuon Production at D0}'',} \textit{ Phys. Rev. Lett.}
  \textbf{ 95} (2005) 161602,
  \href{http://dx.doi.org/10.1103/PhysRevLett.95.161602}{\doi{10.1103/PhysRevLett.95.161602}}.

\bibitem{Chatrchyan:2011jx}
\hrefCMSnoop {} {{ CMS} Collaboration, ``Search for large extra dimensions in
  the diphoton final state at the {L}arge {H}adron {C}ollider'',} \textit{
  JHEP} \textbf{ 05} (2011) 85,
  \href{http://dx.doi.org/10.1007/JHEP05(2011)085}{\doi{10.1007/JHEP05(2011)085}}.

\bibitem{cmspaper11:ADDgammagamma}
\hrefCMSnoop {} {{ CMS} Collaboration, ``Search for signatures of extra
  dimensions in the diphoton mass spectrum at the {L}arge {H}adron
  {C}ollider'',} (2011). \href{http://www.arXiv.org/abs/1112.0688}{\texttt{
  arXiv:1112.0688}}. accepted by Phys. Rev. Lett.

\bibitem{ATLAS:2011ab}
\hrefCMSnoop {} {{ ATLAS} Collaboration, ``Search for Extra Dimensions Using
  Diphoton Events in 7 {TeV} Proton-Proton Collisions with the {ATLAS}
  Detector'',} (2011). \href{http://www.arXiv.org/abs/1112.2194}{\texttt{
  arXiv:1112.2194}}. submitted to Phys. Lett. B.

\bibitem{cms:2008zzk}
\hrefCMSnoop {} {{ CMS} Collaboration, ``The {CMS} experiment at the {CERN}
  {LHC}'',} \textit{ JINST} \textbf{ 3} (2008) S08004,
\href{http://dx.doi.org/10.1088/1748-0221/3/08/S08004}{\doi{10.1088/1748-0221/3/08/S08004}}.

\bibitem{pas:CMS-PAS-EWK-10-004}
\href {http://cdsweb.cern.ch/record/1279145} {{ CMS} Collaboration,
  ``Measurement of {CMS} Luminosity'',} CMS Physics Analysis Summary
  CMS-PAS-EWK-10-004, (2010).

\bibitem{Skands08:Pythia8}
\hrefCMSnoop {} {T.~Sj{\"o}strand, S.~Mrenna, and P.~Skands, ``A Brief
  Introduction to {PYTHIA} 8.1'',} \textit{ Comput. Phys. Commun.} \textbf{
  178} (2008) 852,
  \href{http://dx.doi.org/10.1016/j.cpc.2008.01.036}{\doi{10.1016/j.cpc.2008.01.036}}.

\bibitem{Ask09:ADDpythia8}
\hrefCMSnoop {} {S.~Ask {et~al.}, ``Real emission and virtual exchange of
  gravitons and unparticles in \textsc{pythia}8'',} \textit{ Comp. Phys. Comm.}
  \textbf{ 181} (2009) 1593,
  \href{http://dx.doi.org/10.1016/j.cpc.2010.05.013}{\doi{10.1016/j.cpc.2010.05.013}}.

\bibitem{Martin:2009MSTW}
\hrefCMSnoop {} {A.~D. Martin {et~al.}, ``Parton distributions for the
  {LHC}'',} \textit{ Eur. Phys. J. C} \textbf{ 63} (2009) 189,
  \href{http://dx.doi.org/10.1140/epjc/s10052-009-1072-5}{\doi{10.1140/epjc/s10052-009-1072-5}}.

\bibitem{Agostinelli:2002hh}
\hrefCMSnoop {} {{ GEANT4} Collaboration, ``{GEANT4}--a simulation toolkit'',}
  \textit{ Nucl. Instrum. Meth.} \textbf{ A506} (2003) 250,
  \href{http://dx.doi.org/10.1016/S0168-9002(03)01368-8}{\doi{10.1016/S0168-9002(03)01368-8}}.

\bibitem{Frixione02:MCatNLO}
\hrefCMSnoop {} {S.~Frixione and B.~Webber, ``{M}atching {NLO} {QCD}
  Computations and Parton Shower Simulations'',} \textit{ JHEP} \textbf{ 06}
  (2002) 29,
  \href{http://dx.doi.org/10.1088/1126-6708/2002/06/029}{\doi{10.1088/1126-6708/2002/06/029}}.

\bibitem{Frixione10:MCatNLO}
\hrefCMSnoop {} {S.~Frixione {et~al.}, ``{NLO} {QCD} Corrections in
  {H}erwig{++} with {MC}$@${NLO}'',} \textit{ JHEP} \textbf{ 01} (2011) 53,
  \href{http://dx.doi.org/10.1007/JHEP01(2011)053}{\doi{10.1007/JHEP01(2011)053}}.

\bibitem{Nadolsky:2008zw}
\hrefCMSnoop {} {P.~M. Nadolsky {et~al.}, ``Implications of {CTEQ} global
  analysis for collider observables'',} \textit{ Phys. Rev. D} \textbf{ 78}
  (2008) 013004,
  \href{http://dx.doi.org/10.1103/PhysRevD.78.013004}{\doi{10.1103/PhysRevD.78.013004}}.

\bibitem{Corcella:2000Herwig}
\hrefCMSnoop {} {G.~Corcella {et~al.}, ``{HERWIG} 6.5: an event generator for
  hadron emission reactions with interfering gluons (including supersymmetric
  processes)'',} \textit{ JHEP} \textbf{ 01} (2001) 10,
  \href{http://dx.doi.org/10.1088/1126-6708/2001/01/010}{\doi{10.1088/1126-6708/2001/01/010}}.

\bibitem{Golonka:2005Photos}
\hrefCMSnoop {} {P.~Golonka and Z.~Was, ``{PHOTOS} {M}onte {C}arlo: a precision
  tool for {QED} corrections in {Z} and {W} decays'',} \textit{ Eur. Phys. J.
  C} \textbf{ 45} (2006) 97,
  \href{http://dx.doi.org/10.1140/epjc/s2005-02396-4}{\doi{10.1140/epjc/s2005-02396-4}}.

\bibitem{Butterworth:1996Jimmy}
\hrefCMSnoop {} {J.~M. Butterworth, J.~R. Forshaw, and M.~H. Seymour,
  ``Multiparton interactions in photoproduction at {HERA}'',} \textit{ Z. Phys.
  C} \textbf{ 72} (1996) 637,
  \href{http://dx.doi.org/10.1007/s002880050286}{\doi{10.1007/s002880050286}}.

\bibitem{Ballossini08:DrellYanNLO}
\hrefCMSnoop {} {G.~Balossini {et~al.}, ``{E}lectroweak and {QCD} corrections
  to {D}rell{-}{Y}an processes'',} \textit{ Acta Phys. Polon. B} \textbf{ 39}
  (2008) 1675, \href{http://www.arXiv.org/abs/0805.1129}{\texttt{
  arXiv:0805.1129}}.

\bibitem{Carloni07:Horace}
\hrefCMSnoop {} {C.~M. Carloni~Calame {et~al.}, ``{P}recision Electroweak
  Calculation of the Production of a High Transverse{-}Momentum Lepton Pair at
  Hadron Colliders'',} \textit{ JHEP} \textbf{ 10} (2007) 109,
  \href{http://dx.doi.org/10.1088/1126-6708/2007/10/109}{\doi{10.1088/1126-6708/2007/10/109}}.

\bibitem{petriello10:FEWZ}
R.~Gavin\hrefCMSnoop {} { {et~al.}, ``{FEWZ} 2.0: {A} Code for Hadronic {Z}
  Production at Next-to-Next-to-Leading Order'',} \textit{ Comp. Phys. Comm.}
  \textbf{ 182} (2011) 2388,
  \href{http://dx.doi.org/10.1016/j.cpc.2011.06.008}{\doi{10.1016/j.cpc.2011.06.008}}.

\bibitem{Sjostrand:2006za}
\hrefCMSnoop {} {T.~Sj{\"o}strand, S.~Mrenna, and P.~Z. Skands, ``{PYTHIA} 6.4
  Physics and Manual'',} \textit{ JHEP} \textbf{ 05} (2006) 26,
  \href{http://dx.doi.org/10.1088/1126-6708/2006/05/026}{\doi{10.1088/1126-6708/2006/05/026}}.

\bibitem{Ball:2009NNPDF}
\hrefCMSnoop {} {{ NNPDF} Collaboration, ``Fitting Parton Distribution Data
  with Multiplicative Normalization Uncertainties'',} \textit{ JHEP} \textbf{
  05} (2010) 75,
  \href{http://dx.doi.org/10.1007/JHEP05(2010)075}{\doi{10.1007/JHEP05(2010)075}}.

\bibitem{Alekhin:2011PDF4LHC}
\hrefCMSnoop {} {S.~Alekhin {et~al.}, ``{The PDF4LHC Working Group Interim
  Report}'',} (2011). \href{http://www.arXiv.org/abs/1101.0536}{\texttt{
  arXiv:1101.0536}}.

\bibitem{Tricoli:2005nx}
\hrefCMSnoop {} {A.~Tricoli, A.~Cooper-Sarkar, and C.~Gwenlan, ``Uncertainties
  on {W} and {Z} Production at the {LHC}'',} (2005).
  \href{http://www.arXiv.org/abs/hep-ex/0509002}{\texttt{
  arXiv:hep-ex/0509002}}.

\bibitem{Alwall:2011uj}
\hrefCMSnoop {} {J.~Alwall {et~al.}, ``{MadGraph} 5 : going beyond'',} \textit{
  JHEP} \textbf{ 06} (2011) 128,
  \href{http://dx.doi.org/10.1007/JHEP06(2011)128}{\doi{10.1007/JHEP06(2011)128}}.

\bibitem{:2011nx}
\hrefCMSnoop {} {{ CMS} Collaboration, ``Measurement of the inclusive {W} and
  {Z} production cross sections in pp collisions at {$\sqrt{s} = 7\TeV$} with
  the {CMS} experiment'',} \textit{ JHEP} \textbf{ 10} (2011) 132,
  \href{http://dx.doi.org/10.1007/JHEP10(2011)132}{\doi{10.1007/JHEP10(2011)132}}.

\bibitem{Read:2000ru}
\href {http://cdsweb.cern.ch/record/411537} {A.~L. Read, ``Modified frequentist
  analysis of search results (the {CL(s)} method)'',} in \textit{ Workshop on
  Confidence Limits}, p.~81.
\newblock Geneva, Switzerland, 2000.

\bibitem{Cowan:2010js}
\hrefCMSnoop {} {G.~Cowan {et~al.}, ``Asymptotic formulae for likelihood-based
  tests of new physics'',} \textit{ Eur. Phys. J. C} \textbf{ 71} (2011) 1554,
  \href{http://dx.doi.org/10.1140/epjc/s10052-011-1554-0}{\doi{10.1140/epjc/s10052-011-1554-0}}.

\bibitem{RooStats:Ref1}
\hrefCMSnoop {} {L.~Moneta {et~al.}, ``{T}he {R}oo{S}tats {P}roject'',}
  \textit{ {ACAT}2010 {C}onference {P}roceedings} (2010)
  \href{http://www.arXiv.org/abs/1009.1003}{\texttt{ arXiv:1009.1003}}.

\bibitem{Kumar:2006id}
\hrefCMSnoop {} {M.~C. Kumar, P.~Mathews, and V.~Ravindran, ``{PDF} and scale
  uncertainties of various {DY} distributions in {ADD} and {RS} models at
  hadron colliders'',} \textit{ Eur. Phys. J. C} \textbf{ 49} (2007) 599,
  \href{http://dx.doi.org/10.1140/epjc/s10052-006-0054-0}{\doi{10.1140/epjc/s10052-006-0054-0}}.

\bibitem{Mathews:2004xp}
\hrefCMSnoop {} {P.~Mathews {et~al.}, ``Next-to-Leading order {QCD} corrections
  to the {Drell-Yan} cross section in models of {TeV}-scale gravity'',}
  \textit{ Nucl. Phys. B} \textbf{ 713} (2005) 333,
  \href{http://dx.doi.org/10.1016/j.nuclphysb.2005.01.051}{\doi{10.1016/j.nuclphysb.2005.01.051}}.

\end{thebibliography}\endgroup

\cleardoublepage \appendix\section{The CMS Collaboration \label{app:collab}}\begin{sloppypar}\hyphenpenalty=5000\widowpenalty=500\clubpenalty=5000\textbf{Yerevan Physics Institute,  Yerevan,  Armenia}\\*[0pt]
S.~Chatrchyan, V.~Khachatryan, A.M.~Sirunyan, A.~Tumasyan
\vskip\cmsinstskip
\textbf{Institut f\"{u}r Hochenergiephysik der OeAW,  Wien,  Austria}\\*[0pt]
W.~Adam, T.~Bergauer, M.~Dragicevic, J.~Er\"{o}, C.~Fabjan, M.~Friedl, R.~Fr\"{u}hwirth, V.M.~Ghete, J.~Hammer\cmsAuthorMark{1}, M.~Hoch, N.~H\"{o}rmann, J.~Hrubec, M.~Jeitler, W.~Kiesenhofer, M.~Krammer, D.~Liko, I.~Mikulec, M.~Pernicka$^{\textrm{\dag}}$, B.~Rahbaran, C.~Rohringer, H.~Rohringer, R.~Sch\"{o}fbeck, J.~Strauss, A.~Taurok, F.~Teischinger, P.~Wagner, W.~Waltenberger, G.~Walzel, E.~Widl, C.-E.~Wulz
% ...
\vskip\cmsinstskip
\dag:~Deceased\\
1:~~Also at CERN, European Organization for Nuclear Research, Geneva, Switzerland\\
% ...

\end{sloppypar}
\end{document}